\documentclass{sig-alternate-10pt}
\usepackage{graphicx}
\graphicspath{{fig/}}
\begin{document}
%
\conferenceinfo{IMC}{'15 Tokyo, Japan}

\title{A new intrinsic way to measure IXP performance: \\ an experience in Bolivia}

%
%
%
%
%

\numberofauthors{3} 
%
\author{
%
\alignauthor Esteban Carisimo\\
       \affaddr{Facultad de Ingenier\'ia, Universidad de Buenos Aires}\\
       \affaddr{Paseo Col\'on 850-C1063ACV Buenos Aires, Argentina}\\
       \email{carisimo@cnet.fi.uba.ar}\\
\alignauthor Hernan Galperin\titlenote{CONICET}\\
       \affaddr{Universidad de San Andr\'es}\\
       \affaddr{Vito Dumas 284, Buenos Aires, Argentina}\\
       \email{hgalperin@udesa.edu.ar}\\
\alignauthor J.I. Alvarez-Hamelin\titlenote{INTECIN (U.B.A-CONICET)}\\
       \affaddr{Facultad de Ingenier\'ia, Universidad de Buenos Aires}\\
       \affaddr{Paseo Col\'on 850-C1063ACV Buenos Aires, Argentina}\\
       \email{ihameli@cnet.fi.uba.ar}
}
%
\date{6 May 2015}

\maketitle
\begin{abstract}

Bolivia, a landlocked emerging country in South America, has one of the smallest networks in the whole Internet. Before the IXP implementation, delivering packets between national ISPs had to be sent them through international transit links. Being aware of this situation and looking for increasing the number of users, Bolivian government enacted a law to gather all national ISPs on a single IXP in 2013.

In spite of several articles have researched about this topic, no one before has set the focus on measuring the evolution of end-users parameters in a South American developing country, moreover  after a significant changing on the topology. For the current work, we have mainly studied hop, latency, traffic and route variation, a long a seven months. Topology have not been studied because Bolivian ISPs must be connected each others under legal obligation.

To achieve our measurement goals, and under absence of global-scale measuring projects in this country, we have developed our own active-measurement platform among local ASes. During the platform development we had to deal with local ISP fears, governmental agencies and regulation pressures.  

We also survey the main previous papers on IXP analysis, and we classfied them on obtained data and their sources.

\end{abstract}




\section{Introduction}
  Since the mid-1990s Internet eXchange Points (IXPs) have proliferated worldwide. The rapid growth of this architecture has called the attention of the networking research community, with several published articles in recent years (e.g.,  \cite{ager2012anatomy,cardona2012history,chatzis2013there,lodhi2014using,augustin2009ixps,xu2004properties}).
  
  IXP's aim can be summarized in this {\em leitmotiv} taken from \cite{chatzis2013there}: ``exchange local traffic locally''.
  This means that local traffic will be interchanged using local links instead of international transit networks ~\cite{dhamdhere2010internet,gregori2011impact}.
  Switching to local networks to deliver local traffic has important technical and economic advantages.
  Technical advantages include the reduction in {\em Round  Trip Time} and {\em Number of hops} to reach destinations.
  Economic advantages include the ability to reach several local ASes with a single local link and savings from reduced usage of international transit links.
  
  In this paper we examine the implementation of a new IXP in Bolivia. As a landlocked low-income country, Bolivia has one of the smallest Internet networks in Latin America. 
  As part of its efforts to promote Internet growth, in October 2012 the Bolivian government enacted a law requiring that ISPs with international connectivity exchange traffic locally through a single interconnection point.
  The goal was to reduce costs and increase network performance, thus promoting infrastructure investments and adoption. In November 2013, the six largest ISPs (AXS, COTAS, ENTEL, COMTECO, NUEVATEL and TELECEL) started exchanging traffic at Bolivian IXP, as the ISP came to be known.
 
  For research purposes, the case of Bolivian IXP has some desirable characteristics. First, Bolivia ranks at or near the bottom of Internet development (however measured) in Latin America. 
  It therefore presents a fertile environment for testing the impact of a new IXP on network performance and traffic patterns. Second, there are not any previous work who have studied Latin American IXP through active measurements.
  Third, Bolivian IXP is the result of a new interconnection law which requires large ISPs to exchange traffic locally. This is critically different from other cases in which IXP participation is voluntary, and therefore its impact is highly dependent on which networks choose to participate.

  Given our goals and the characteristics of the networking environment in Bolivia, existing tools, resources and techniques used in previous papers to analyze IXP behavior and evolution were deemed not suitable.
  We therefore developed a distributed {\em ad-hoc} measurement platform, with probes in different cities and ASes in Bolivia. 
  Using this platform, we were able to generate a \texttt{traceroute}-based dataset, which was processed to obtain several parameters such as \emph{Round Trip Time} and \emph{Number of hops}. 
  We also defined two new parameters called {\em Local Routes} and {\em Accessible time}. 
  These parameters are analyzed to understand changes in network performance related to the implementation of Bolivian IXP.
  
  
  Using resources external to our platform (not public) we have also measured traffic growth and the growth of services provided from Bolivian IP addresses.

  The paper is organized as follows: the next section surveys the state of art in the study of IXPs; Section~\ref{pladmed} describes the methodology used in our measurement platform; Section~\ref{Results} discusses the results obtained, while Section~\ref{conclusion} presents conclusions and implications for further work.
  
 \section{Related Work}~\label{rw}
The IXP literature can be organized according to the measurement techniques used to generate datasets. These can be classified in six groups: (i) Using internal vantage points, (ii) Analyzing IXP's website, (iii) Mining into Public Databases, (iv) ICMP-based measurements, (v) BGP-based measurements and (vi) Developing platforms using public resources. However, several researches have applied several techniques to achieve their goals. For this reason, a long this introduction we will match the most relevant literature with the categories before mentioned. As result of this classification we will obtain a table which will set a milestone in IXP literature classification.

\begin{table*}[t]
\centering
\footnotesize
\begin{tabular}{|p{2.4cm}||p{2.6cm}|p{2.6cm}|p{2.6cm}|p{2.6cm}|p{2.6cm}|}
\hline
 & {\bf Int/Ext VG} & {\bf Web} & {\bf Public DB}  & {\bf BGP} &  {\bf traceroute}  \\
\hline\hline
{\em AS classification} & {\bf Int:} Ager \emph{et al.}\cite{ager2012anatomy} & Cardona {\em et al.}\cite{cardona2012history} 
                                                                                                                           Augustin \emph{et al.}~\cite{augustin2009ixps} & Augustin \emph{et al.}~\cite{augustin2009ixps} 
                                                                                                                                                                                                          Chatzis \emph{et al.}~\cite{chatzis2013there}  
                                                                                                                                                                                                          Lodhi \emph{et al.}~\cite{lodhi2014using} ~     
                                                                                                                                                                                                           Xu \emph{et al.}~\cite{xu2004properties}        & Augustin \emph{et al.}~\cite{augustin2009ixps} 
                                                                                                                                                                                                                                                                                       Xu \emph{et al.}~\cite{xu2004properties}           & Augustin \emph{et al.}~\cite{augustin2009ixps}
                                                                                                                                                                                                                                                                                                                                                                      Xu \emph{et al.}~\cite{xu2004properties}  \\ \hline
{\em Peering matrix}    &                                                                                 & Cardona {\em et al.}\cite{cardona2012history} 
                                                                                                                             Augustin \emph{et al.}~\cite{augustin2009ixps} 
                                                                                                                             Lodhi \emph{et al.}~\cite{lodhi2014using}         & Augustin \emph{et al.}~\cite{augustin2009ixps} 
                                                                                                                                                                                                           Chatzis \emph{et al.}~\cite{chatzis2013there}  
                                                                                                                                                                                                           Lodhi \emph{et al.}~\cite{lodhi2014using} ~     
                                                                                                                                                                                                           Xu \emph{et al.}~\cite{xu2004properties}         & Ager \emph{et al.}\cite{ager2012anatomy} ~~ 
                                                                                                                                                                                                                                                                                         Augustin \emph{et al.}~\cite{augustin2009ixps} 
                                                                                                                                                                                                                                                                                         Xu \emph{et al.}~\cite{xu2004properties} Gupta {\em et al.}~\cite{gupta2014peering}
                                                                                                                                                                                                                                                                                                                                                    & Ager \emph{et al.}\cite{ager2012anatomy} Augustin \emph{et al.}~\cite{augustin2009ixps}  ~ Xu \emph{et al.}~\cite{xu2004properties} Gupta~{\em et al.}~\cite{gupta2014peering} Fanou~{\em et al.}\cite{fanou2015diversity} This~work \\ \hline
{\em Traffic}                 & {\bf Int:} Ager \emph{et al.}\cite{ager2012anatomy} 
                                          {\bf Int:} This work                                             & Cardona {\em et al.}\cite{cardona2012history} 
                                                                                                                        Chatzis \emph{et al.}~\cite{chatzis2013there}   &   Lodhi \emph{et al.}~\cite{lodhi2014using}   & & \\ \hline
{\em IXP users QoS}  &                                                                                  & & & & Gupta {et. al.}~\cite{gupta2014peering} This work \\ \hline
{\em Services}             & {\bf Int:} Ager \emph{et al.}\cite{ager2012anatomy} 
                                        {\bf Ext:} This~work                                                                  & & & & \\ \hline
\end{tabular}
\caption{Taxonomy based on IXP papers, as function of sources (columns) and measured parameters (rows). The first ones are: Internal or External Vantage Points, IXPs' Website, Public Databases, BGP routing tables, traceroute-based tools. The second ones are: AS classification (geographic, size, Tier, business type), Peering Matrix (ASes relationship, local routes, available time), IXP users QoS (distance, RTT, inter-hop RTT-difference), Traffic (amount, classification, port utilization), Services (classification, amount).}\label{taxo_ixps} 
\end{table*}
 
 \paragraph{Using internal vantage points} This category is represented by the work of Ager \emph{et al.}~\cite{ager2012anatomy}. This paper is focused on describing the main characteristics of one of the largest European IXPs. For this purpose the authors installed a device inside the IXP which sampled the packets that passed through it. This was accomplished using sFlow which was developed and configured for sampling just the headers of one out of 16k packets. Layer-2, IP and TCP/UDP headers provide enough information to achieve these goals while keeping the privacy of the packets passing through the IXP.
 
 sFlow allowed the authors to identify how many ASes were connected at the IXP and classified them by tier, country and continent. Ager \emph{et al.} also processed the headers and calculated the packet rate, bandwidth and daily average volume of traffic.
 
 The most relevant finding from this tool was the ability to obtain the peering matrix. It is very usual that AS-level topology graphs are incomplete, mainly because peering links between ASes are hidden. Ager \emph{et al.} introduced around 50k new links which are present at this IXP. Their work shows that IXPs generate large amounts of new peering links, thus enhancing AS-level topology graphs.
 
 \paragraph{Analyzing IXP's website} Most IXPs, in particular large ones, publish data about their traffic and performance. Some examples are AMS-IX~\cite{amsix}, DE-CIX~\cite{decix} and Slovak-IX (SIX)~\cite{six}, which Cardona \emph{et al.}~\cite{cardona2012history} study in their paper.
 
 These authors use snapshots taken from SIX's website since 1997. The snaphots were available from \texttt{wayback machine}~\cite{wayback}, a project which attempts to create an archive of Internet websites by periodically saving webpages on their own database.
 
 Using this technique, Cardona \emph{et al.} analyzed parameters such as the peering matrix, the number of peering links through the IXP and their evolution. This paper also analyzed traffic growth using the \texttt{mrtg} graph published on the SIX  website. Cardona {\em et al.} also take \emph{raw} data to identify inbound and outbound traffic, and classified each category depending on which type of AS had generated the traffic. Another parameter typically available from IXP websites is the number of ASes connected at the IXP, which Cardona \emph{et al.} also analyze in their work.
 
 This technique is also used by Augustin \emph{et al.}~\cite{augustin2009ixps}, though this paper also takes advantage of another measurement technique discussed below. 
 Lodhi \emph{et al.}~\cite{lodhi2014using} and Chatzis \emph{et al.}~\cite{chatzis2013there} have also checked their results against website information.
 
 \paragraph{Mining public databases} There are two well-known databases for networking researchers to obtain information about IXPs: \textsc{PeeringDB}~\cite{pdb} and \textsc{Packet Clearing House (PCH)}~\cite{pch}. These databases store information about ASes and their peers, so that researcher are able to obtain peering links and IXP members.
 
 Several articles have used these sources, though Chatzis \emph{et al.}~\cite{chatzis2013there} have criticized the accuracy of IXP records in these databases. First of all, \textsc{PCH} and \textsc{PeeringDB} will always be incomplete since data availability and input is voluntary. Moreover, PCH has a policy to never drop an IXP from its list even when IXPs are not operational.
 
 Another paper that uses these databases is Lodhi \emph{et al.}~\cite{lodhi2014using}, which uses \textsc{PeeringDB} to generate its dataset. In this paper the focus is not on the IXPs themselves but rather on the Peering Ecosystem (relationships between ASes), where the IXPs are a subset of it. This paper also criticizes the accuracy of the available IXP databases. Their main argument is that small IXPs and those from developing regions are not accurately represented in these databases. 
 
 Lodhi \emph{et al.} have attempted to determine how accurate IXP records are in \textsc{PeeringDB}. For this purpose they compared each of the top-20 IXPs present in \textsc{PeeringDB} with data published in the IXPs own websites.  Other parameters that appear in Lodhi \emph{et al.} are: presence of CDNs at IXP (i.e., Google or Netflix) and rate of links from ASes to IXPs, based on a stable group of ASes. Augustin \emph{et al.}~\cite{augustin2009ixps} have also completed their dataset mining into public databases.

 
 \paragraph{ICMP-based measurements} \texttt{Ping} and \texttt{traceroutes} are the most common probes in this group of papers. Some world-scale measurement projects such as Ark CAIDA~\cite{caida} or DIMES~\cite{dimes} have generated dataset with information provided by \texttt{traceroutes}. Public availability and periodical renewal of these data make them very useful, especially for the networking research community, which has published several papers based on this information.
 
 In articles where IXP is the main focus, CAIDA and DIMES datasets are widely used although they do not always contain information about IXPs~\cite{oliveira2010completeness,gregori2011impact}. Depending on where the vantage points of these projects are located, many IXPs will remain hidden. It is impossible to reveal all IXPs using these types of datasets, and for this reason other techniques are used to analyze IXP performance. Among the papers that have used these datasets are Augustin \emph{et al.}~\cite{augustin2009ixps} and Xu \emph{et al.}~\cite{xu2004properties}.
 
 Gupta {\em et al.}~\cite{gupta2014peering} has developed their own ICMP-based platform using {\em BISmark}
 
 \paragraph{BGP-based measurements} This technique is based on BGP-dumped tables, which can be processed to acquire some parameters about network topology.
 
 
 There are many projects that periodicallygather BGP tables from border routers. The most used by the networking research community are RouteViews~\cite{views2000university} and RIPE RIS~\cite{ripe2006routing}. These have been used by many researches, but they are not entirely accurate~\cite{chang2004towards,dhamdhere2008ten,oliveira2010completeness}. At times, ASes want to keep  some routes private, especially when they do not wish to publish all peering links. Notice that some peering links of the AS, to which the border router belongs to, will be present at this table but not all of them~\cite{he2009lord}. Additionally, peering links of ASes into the tier tree of that AS will be also missing.
 
 
 Papers using BGP tables reveal this technique produces even less information about IXPs than \texttt{traceroutes} datasets~\cite{huffaker2002topology}. Yet it has been used in several articles such as Augustin \emph{et al.} and Xu \emph{et al.}
 
 Xu \emph{et al.} combine passive and other techniques, finding only 82 IXPs out of 144 present in PCH in 2004 (remember that PCH rarely deletes IXPs from its records). Researchers have been improving Xu's techniques for more than 10 years, so today  more accurate results can be achieved. 
 
 Under this label, we should include Hurricane Electric (HE) peering information. HE provides BGP peering information on its webpage, and this information was used by Gupta {\em et al.}~\cite{gupta2014peering} to inferred about African ASes relationship.
 
 \paragraph{Developing platforms using public resources} We have mentioned that \texttt{traceroutes} datasets from world-scale project sometimes do not have information about IXPs. The main reason is the absence of probes in ASes connected to those missing IXP.
 
 Augustin's \emph{et al.}~\cite{augustin2009ixps} goal was to find as many IXPs as possible. For this reason the authors developed a platform to get information about those IXP which do not appear in \texttt{traceroutes}-based or BGP-based datasets. This platform was created by placing Looking Glasses (LG) in ASes which are linked to IXPs. Looking Glass is a service which allows using a server's IP address to run commands such as \texttt{ping} or \texttt{traceroute}. Augustin \emph{et al.} used these LG as vantage points, executing \texttt{traceroutes} to generate a dataset where information about IXP is included.
 
 Combining this technique and others mentioned in previous sections, Augustin \emph{et al.} have found 58k peering links through IXPs. With this information the authors examined parameters such as the ranking of IXPs by number of peering links and compared this ranking with different datasets. They have also created a ranking of IXP by number of members. 

\section{Gathering data}~\label{pladmed}

The main goal of this paper is to analyze the evolution of end-users parameters on the first Bolivian IXP. 
Using these mentioned metrics we could analyze improvements on Internet performance in the country, which probably set a background for enlarging the number of user in the following years. Gupta {\em et al.}~\cite{gupta2014peering} has recently used end-users latency also in low-income countries. 
Further, we will explain deeper how this article has used it.
We note that public databases contain almost no information on Bolivian ASes or the new IXP.


For example there are no BGP tables that contain information about the Bolivian IXP neither RouteViews~\cite{views2000university} nor RIPE RIS~\cite{ripe2006routing}. The absence of information about emerging countries in these databases was already mentioned at Gupta {\em et al.}.

In addition \texttt{Traceroute} measurement projects like Ark-CAIDA~\cite{ark} or DIMES~\cite{dimes} do not have probes in Bolivia, so these \texttt{traceroute} datasets are of little value for our goals. For example, in Figure~\label{caida_bo} we have checked the Bolivian ASes map using ``I'm here!''~\cite{BAHB2008} \\(http://http://lanet-vi.fi.uba.ar/i\_am\_here\_/buscar.py), which uses CAIDA as data source, and we have not found a structure that could match with the Bolivian IXP.

\begin{figure}[t]
\includegraphics[width=\columnwidth]{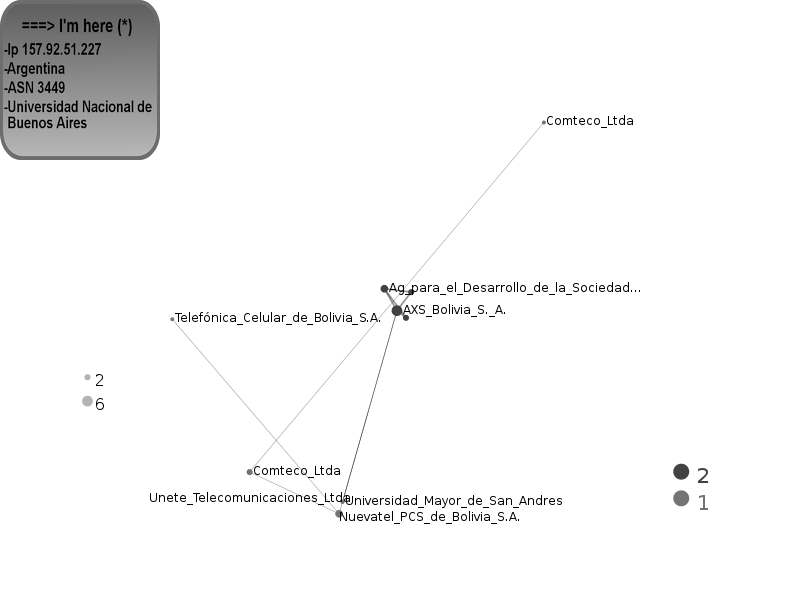}
\caption{This graph was produced using ``I'm here!'' and CAIDA's dataset. As it is shown, this graph does not look like a full mesh, which would be expected for Bolivian IXP peering policies. This graph must be seen carefully because some nodes might be overlapped, hiding the below ones.
}\label{caida_bo}
\end{figure}

We were unable to perform sampling techniques such as sFlow because they require permissions that network administrators in Bolivia were unwilling to grant. Further, local ISPs refused the installation of any type of vantage point from within the IXP, citing security and privacy concerns.

Another alternative would have been to develop a LG platform with probes placed in Bolivia. However, as of now there are no LG probes installed on Bolivian ASes.

According to our goals, QoS tools, such as SamKnows\footnote{SamKnows: QoS measurement tool. https://www.samknows.com/},  did not cover most of our pretended parameters. Moreover, this well-known system require a local server which needs at least 1 Gbps. This requirement  was impossible to achieve using the Bolivian network.

In addition to absence of technical resources we had to deal with political, social and legal issues.
Bolivian government and the national regulator (ATT~\cite{att}) were interested in auditing if the law was being fulfilled.
In the opposite side, ISPs were afraid of government would punish them if their services did not fit tightly to the telecommunication law.
Moreover, ENTEL, who is the largest Bolivian ISP and state-owned telecommunication company, was chosen to hold Bolivian IXP, giving to it addresses and infrastructure.
This situation drove to the ISPs to refuse of any kind of device or sharing data because they felt threatened.
Finally, ATT and we agreed to develop a co-owned platform to satisfied our researching goals and their regulation necessities.

Other alternatives that we discard were deploy the platform using RIPE Atlas~\cite{atlas} or Planet-Lab\cite{planetlab} probes. 
On one hand, RIPE Atlas would have been a suitable option, and there are recently information in \cite{fanou2015diversity} about it performance in Africa, a similar scenario as Bolivia. However, this platform would have not given us enough flexibility to edit scripts and network measurement parameters. 
On the other hand Planet-Lab would have provided us flexibility required, however, there was not Planet-Lab resources in Bolivia, and we did not find any partner who would want connect a probe in its network.

As a result we opted for developing a distributed measurement platform, called PladMeD (Spanish acronym for \textbf{Pla}taforma \textbf{d}e \textbf{Me}dici\'on \textbf{D}is-tribuida) to obtain data from the network's edges. 


PladMeD is a set of probes connected to Bolivian ASes. These probes run\texttt{traceroutes}-like tools to gather data. Our goal was to deploy 12 probes in the 6 ISPs that are connected to Bolivian IXP. As of June 2014 we were able to install 3 probes in the two largest cities: two in La Paz (one of which is not operational as of January 2015) and one in Santa Cruz de la Sierra. Further deployments are expected throughout 2015.

Dall'Asta {\em et al.}~\cite{DAHBVV2006} shows that having at least two probes, some properties can be obtained with a high degree of confidence. This processing is not enough to obtain information about all links. Indeed, invisible links are those that carry traffic only certain ISPs, i.e., the P2P routes.

Choosing the appropriate hardware and software for the probes probes was a key concern. We decided to use {\bf Raspberry Pi}~\cite{rpi} model B as probes. We chose this hardware due to its low price, low-power consumption, and its ability to run Linux and be connected to the Internet through a RJ-45 connector. Each Raspberry Pi on PladMeD runs Raspian OS~\cite{raspbian}, and generates \texttt{traceroutes} measurements by running \texttt{scamper}~\cite{luckie2010scamper} and storing data with \texttt{MySQL}~\cite{mysql2001mysql}. Moreover, \texttt{scamper} is configured to send \texttt{Paris-traceroute}~\cite{augustin2006avoiding} probes and thus avoid wrong paths.

For this projects we have used a Raspberry-Pi platform instead of a BISmark~\footnote{BISmark:  Broadband Internet Service Benchmark or BISmark. http://projectbismark.net/}, such as Gupta {\em et al.}, because we thought that Raspberry-Pi devices suit better. 
Although BISmark is able to run on Raspberry Pis, this software release was after we started our measurements in Bolivia (June 2014).

The key differences between PladMeD and other existing measurement tools and datasets are as follows. First, PladMeD was built as a \emph{ad-hoc} distributed platform developed specifically to study IXPs. Second, PladMeD has placed probes in Bolivian ASes. A similar functionality could be obtained by LG. However, as noted there are no LG servers in Bolivia. PladMeD's probes are installed in residential connections, and as such measurements are obtained from the network's edges as opposed to from the network's core. 





One of our key goals was to examine the relevance of the new IXP in the exchange of local traffic, that is when source and destination address are located in Bolivia. As such we needed to generate traffic where the local source point was PladMeD and the local destination was a Bolivian IP address. We took from LACNIC~\cite{lacnic} public information about which IPv4 netblocks were given by this RIR~\footnote{RIR: Regional Internet registry} to Bolivia. We then split each IPv4 netblock using CAIDA's technique called ``IPv4 Routed /24 Topology''~\cite{claffy2009internet}. From each resultant network we chose a destination address as target for our \texttt{traceroute}.

We have been executing \texttt{traceroutes} 20 hours per day since the deployment of the first probe in June 2014. Although PladMeD's Raspberries are capable of processing data, we only use PladMeD probes as vantage points. The Raspberry Pis send daily data generated to a central server which is placed outside Bolivia. Moreover, this server receives information from all the probes, which allows us to have a full perspective of the IXP.

Yet some small but important processes are performed at the probe. After \texttt{scamper} ends its execution, a \texttt{Python} routine takes \emph{raw} the data and stores it in a \texttt{MySQL} database. This Python script also makes two important checks: first it verifies if the \texttt{traceroute} contains (or does not contain) an address allocated to the Bolivian IXP subnetwork. We are able to execute this check because ATT~\cite{att} has provided us with the IP prefix used at the IXP. At Bolivian IXP all border routers reply each ICMP packets,   using IP addresses assigned to the IXP and not other. Second the script checks is if all hops of the \texttt{traceroute} correspond to IP addresses that LACNIC has assigned to Bolivia. If we find a private address (RFC 1918) we do not have enough information about, we consider that it is also located in Bolivia. After these two checks, we classify each \texttt{traceroute} in one of following four categories:

\begin{itemize}
 \item {\bf IXP}: Passed through the IXP and full path has only Bolivian addresses.
 \item {\bf P2P}: Did not passed through the IXP and full path has only Bolivian addresses.
 \item {\bf Internationals}: Did not passed through the IXP and path contains at least one non-Bolivian address.
 \item {\bf Misbehavior}: Passed through the IXP and path contains at least one non-Bolivian address.
\end{itemize}

Using these four categories we classify the routes between two Bolivian ASes obtained through PladMeD. 
When each hop in the route is within the country and packets pass through the IXP, we tag that route as {\bf IXP}. 
If the packet has a complete path with local IP addresses but did not pass though the IXP, it means that the packets were delivered by direct peering links between ASes. This category is called \textbf{P2P}. Since IXPs are installed to keep local traffic local,  if a packet passes through the IXP and then leaves the country we will consider this as a \textbf{Misbehavior} route. Although the Bolivian government has promoted the IXP in order to connect national ASes using local links, we still observe routes between Bolivian ASes that involve international transit networks. Routes that do not contain any IP address of the IXP's subnetwork and have at lest one non-Bolivian address in its path are categorized as {\bf Internationals} routes.

IXP, P2P and Internationals are the three natural categories to analyze improvements after the IXP development. However, we include an extra category called \textbf{Misbehavior}. We had not expected this category before start to run the platform, nevertheless we found some traceroutes which went against the IXP's leitmotiv. We think that this uncommon behavior may be related with a routing misconfiguration, and this category shows how users are affected as consequence of mistakes.

We have already mentioned that Bolivian network development is not too large and given this reason, P2P category might be empty. However, some ISPs already had a few of P2P links previous to the telecommunications law.

National and international classification must be explained carefully. We assumed that netblocks allocated by LACNIC~\cite{lacnic} to Bolivia are those being used in this country. We have not done any geolocalitation check, however, some information could confirm our hypotheses. LACNIC has only assigned to Bolivia approximately 1.3 million addresses, thus it is highly unlikely that Bolivian IPs are being used on other countries. 

In the opposite way that we mention just before, we could find the reverse situation, non-Bolivian IP addresses placed in Bolivia.
However, it does not affect our classification because when we noticed a foreign address, those traceroute contain more than one foreign address, thus that route was classified correctly as International. 
These foreign addresses placed in Bolivia are only used to connect local ASes with international transit networks, they are not used to provide access to end-users, where is our focus.

Although Gupta {\em et al.}~\cite{gupta2014peering} has also classified which routes left Africa and which not, its method is different from our. In African IXPs analysis, they have used end-users latency as a threshold to determinate which routes had to be marked up as Internationals. In the current work, we have analyzed end-users latency only as QoS improvements. Even though we have not used latency as a threshold, we probed classification consistence through median latency expected values.

For the purposes of the study we have used a division of the Bolivian IP addresses space. Even though this division does not perfectly match with the actual network division, it is widely used \cite{claffy2009internet}. In particular this kind of division is useful to measure the \emph{intensity} of the /24 routing prefixes through our four categories. Yet this division can also introduce biases. On one hand, when a probe inside a specific AS performs \texttt{traceroutes} to prefixes belonging to the same AS, these \texttt{traceroutes} will be classified as \textbf{P2P} although technically they should not. On the other hand, when packets are being sent to another AS, it is possible that not all the networks belonging to the destination AS pass through the IXP; therefore we measure the \emph{intensity} of routes through our different categories.

When ASes are connected to an IXP and services are hosted at local servers, local clients of these services will enjoy lower Round Trip Time (RTT). For this reason, the development of the IXP will encourage services provided from Bolivian IP addresses to grow. Based on this assumption, as part of this project we also measure the number of servers in Bolivia running the most popular services. To analyze this we scan several ports in Bolivian networks using \texttt{ZMap}~\cite{durumeric2013zmap}.

{\tt ZMap} scans a certain port for a given network and obtains data about which addresses have this port open. At times, host serves have specific ports even though they are not providing these services. For this reason, we consider an active service when the IP address appears at least in 3 out of the last 5 measurements. Using this method we ensure the inclusion of small servers which could be down during our scan and we discard occasional port opening. We run \texttt{ZMap} on odd days to follow local Internet services development.


\begin{figure}[t]
\includegraphics[width=\columnwidth]{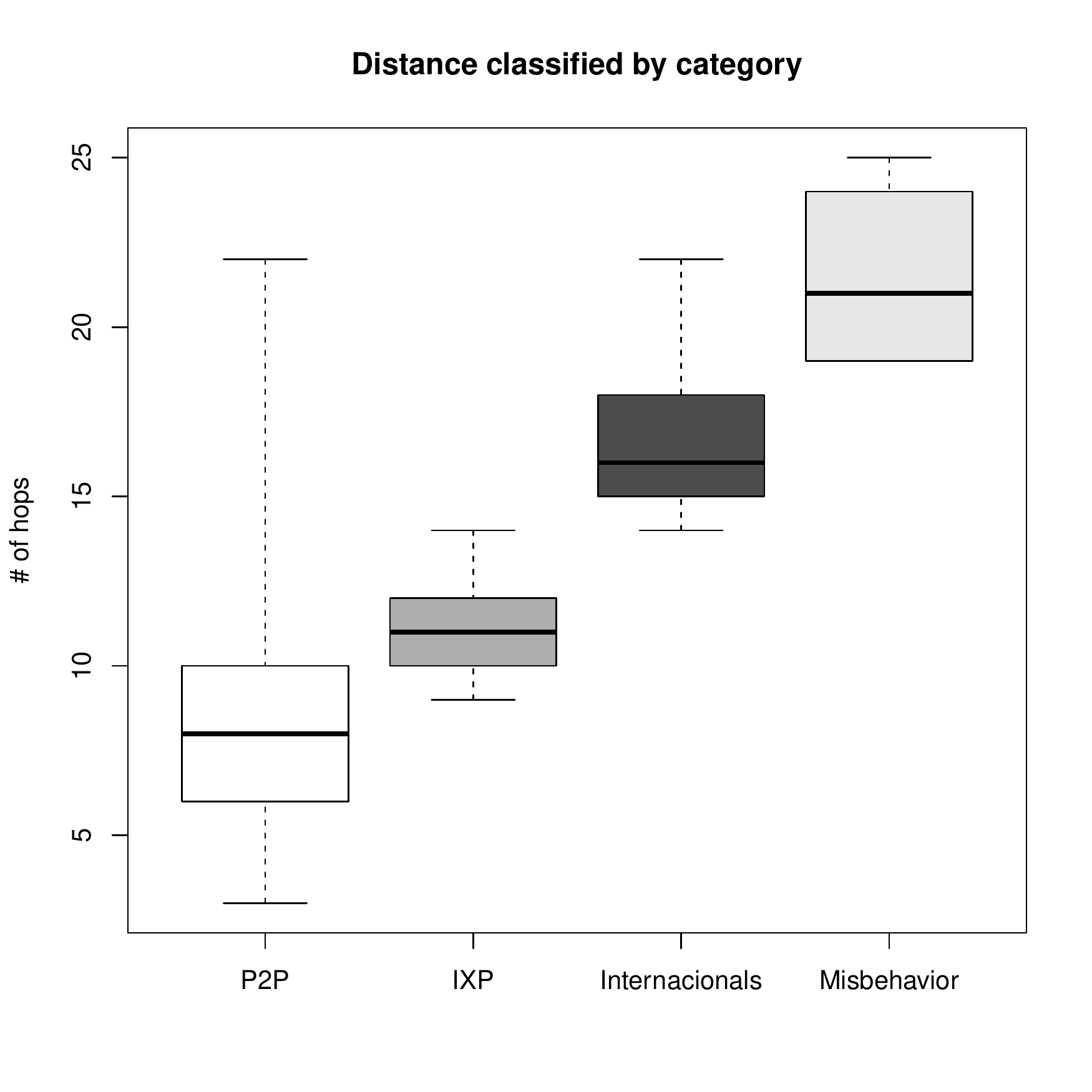}
\caption{Box-plots showing the number of hops for all categories, from 22-Jun to 11-Jan.}\label{resu_hops}
\end{figure}

\begin{figure}[t]
\includegraphics[width=\columnwidth]{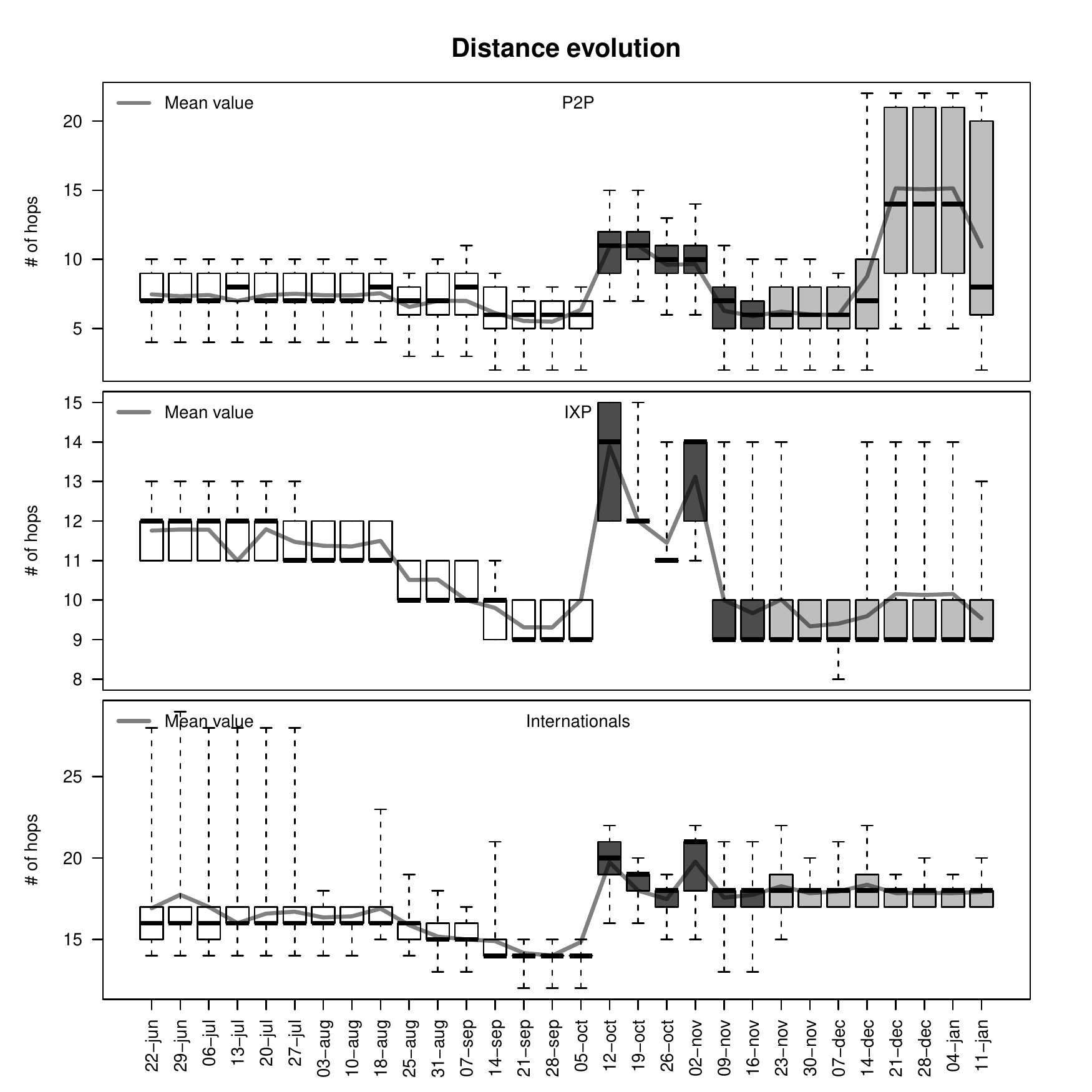}
\caption{Temporal evolution of boxplots displaying the number of hops for main categories. 
The probe in La Paz corresponds to the white boxplots, both probes running at same time are in dark gray, and only Santa Cruz probe period is in gray. Mean is drawn as a grey line.
}\label{resu_hops_t}
\end{figure}

\section{Results}~\label{Results}

In this section we present the results of our measurements between June 2014 and January 2015. 

As explained, we divided Bolivian netblocks to generate smaller networks, and for each one we have selected one IP address. 
We understand that all the addresses of a network are not always used. 
As a result some \texttt{traceroutes} performed to a network will not have an answer from the last hop. 
To enhance the last hop response rate, we selected the target address of each network using \texttt{ZMap's} results. 
If we do not have any active service in a /24 network, we select the target randomly. 
In addition, we have only processed \texttt{traceroutes} that obtain an answer from their target.
Using these parameters we have obtained a last hop response rate of about 50\%.

PladMeD has been active since June, 2014. 
However, its architecture has changed during the reported period, and can be divided into three subperiods:

\begin{enumerate}
 \item from 15-Jun to 12-Oct: only La Paz probe active;
 \item from 12-Oct to 15-Nov: La Paz and Santa Cruz de la Sierra probes active;
 \item from 12-Nov to 11-Jan: only Santa Cruz de Sierra probe active.
\end{enumerate}


The {\em number of hops} is a quantity related to length of the paths. P2P routes are expected to be the shortest ones, followed by IXP routes, and finally by International ones.  Figure~\ref{resu_hops} displays the boxplots (5\% percentile, 1st quartile, median, 3rd quartile and 95\% percentile) of the four types of routes.
Median values are as expected: P2P's median distance is $0.73$ with respect to IXP's median, while International is $1.45$.
The IQR (Inter-Quartile Range, or 3rd quartile minus 1st) is smaller in IXP routes, thus its variance is not as significant; and is larger in International (which is expected) but also in P2P routes. 
This reveals that the internal structure of ISPs is variable in length, probably due to the Bolivia's geography.

\begin{figure}[t]
\includegraphics[width=\columnwidth]{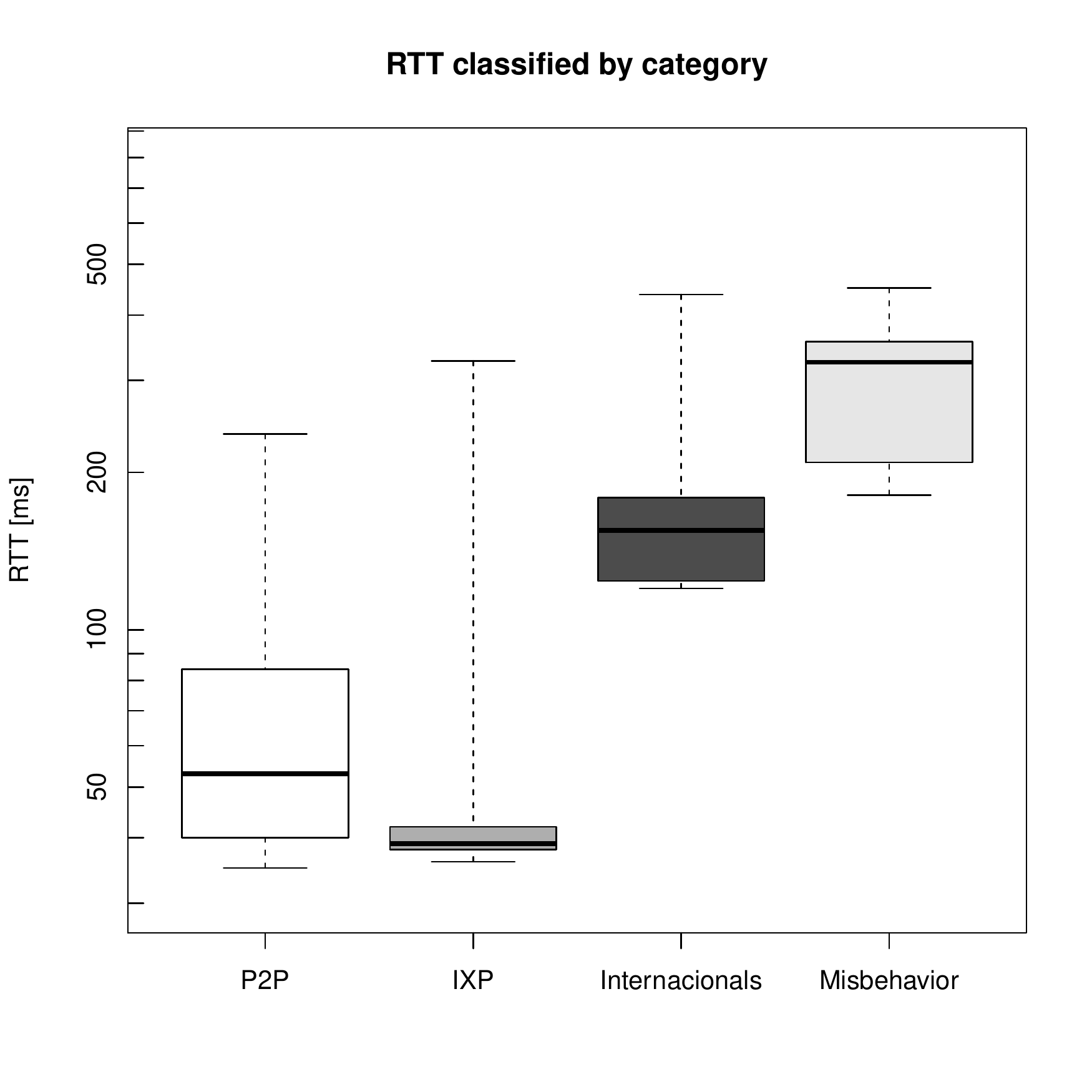}
\caption{Box-plots showing the RTT for all categories, from 22-Jun to 11-Jan. 
The maximum corresponds to the 95 percentile, and minimum to the 5 percentile.
}\label{resu_rtt}
\end{figure}

\begin{figure}[t]
\includegraphics[width=\columnwidth]{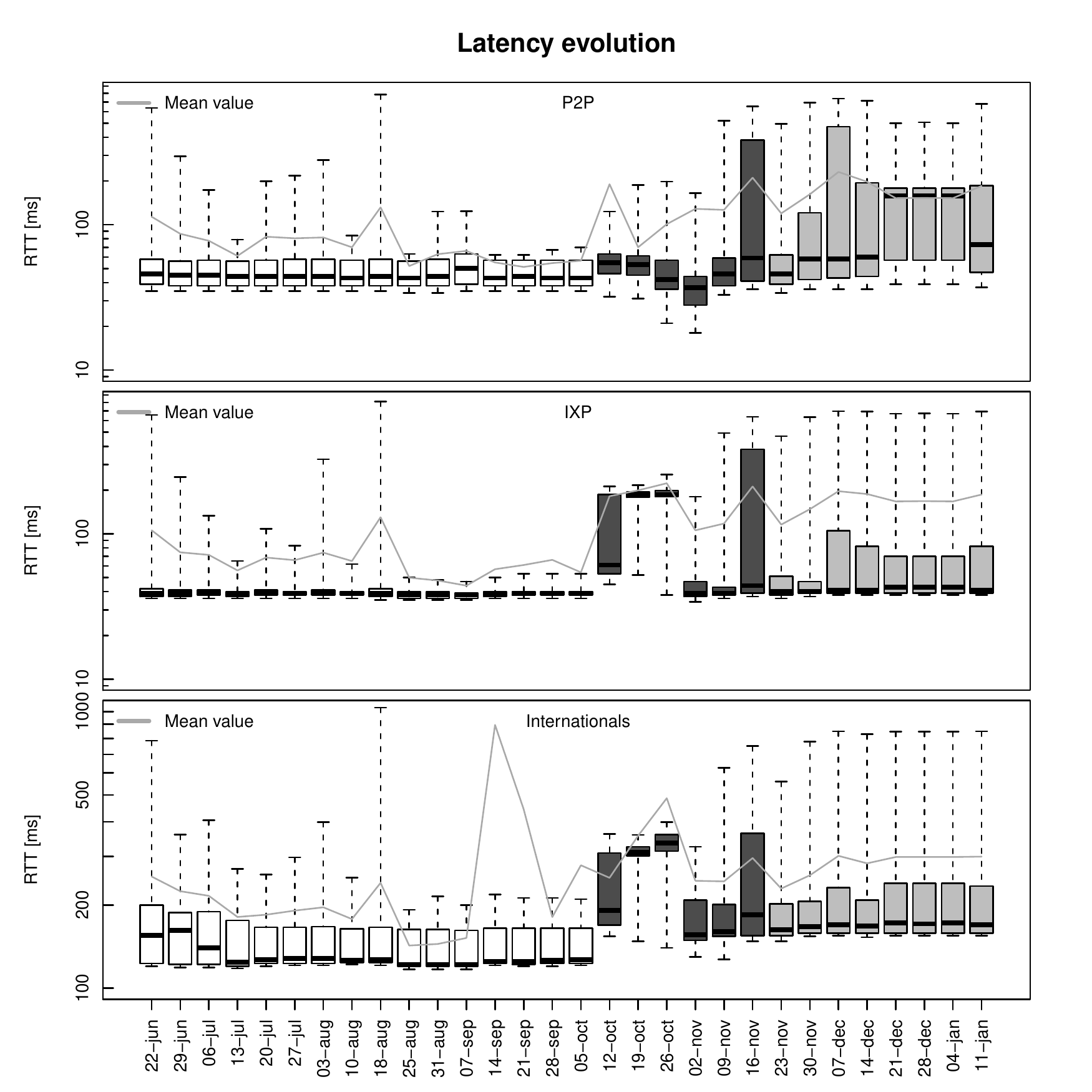}
\caption{Temporal evolution of RTT  for main categories. 
The maximum corresponds to the 95 percentile, and minimum to the 5 percentile.
The probe in La Paz corresponds to the white boxplots, both probes at same time are in dark gray, and only Santa Cruz probe period are those in gray.
}\label{resu_rtt_t}
\end{figure}

Figure~\ref{resu_hops_t} presents the evolution over time in the number of hops for the three main categories (P2P, IXP and International). 
Each box-plot represents a week, and colors represent the different probes: white for La Paz, dark gray when both probes are running, and gray for Santa Cruz; and also their mean values are drawn. 
Over time values tend to stabilize for both probes, but in Santa Cruz they are slightly larger than in La Paz, where the IXP is located. 
In addition our Santa Cruz probe is connected through 4G-LTE network, which naturally increases the number of hops. 
The most important observation is that the tendency in the three curves is similar, confirming the validity of merging the data for both probes.
Mean values are representative in hop counts because they are close to the median, i.e., they have no bias, probably due to hops having a lower upper bound.

Before analyzing {\em Round Trip Time} (RTT) results, a methodological note is needed. 
During our trial period with PladMeD, we noticed RTT results larger than expected. When analyzing many \texttt{traceroutes} in depth we discovered that RTTs increased significantly between the penultimate hop and the target host. 
We present two hypotheses to explain these results. First, we assume that the last hop is connected to a LAN and this local network is in turn connected to the Internet through the penultimate hop. 
LAN networks should not be congested and its delays are typically much smaller than in WAN links.
For these reasons, the RTTs for the last two hops ought to be almost identical.
Our second hypothesis is that this increase in response time is caused by hardware limitations.
We know that Bolivian ISPs use private addresses for household subscribers, and we conclude that NAT~\cite{rfc1631} devices at the last hops must be overloaded, thus artificially increasing \texttt{traceroute} response time.
For this reason, we decided to use the answer time of the penultimate hop as our RTT.

Figure~\ref{resu_rtt} shows the boxplots for all categories (notice that the RTT's axis is in logarithmic scale). 
In this case, the maximum and minimum drawn values represent the 95 and 5 percentiles, respectively.  
The median RTT for P2P routes is $1.36$ larger than for IXP routes, and the median RTT for International routes is $3.97$ larger than in IXP routes. The fact that RTTs in P2P routes are larger is likely related to the link capacity (all links to the IXP have 1Gbps capacity ports, but those of P2P probably have less since they are older). 
The median value in three categories, seems to keep values in the range expected, particularly International. 
The median value is around 130 ms  in International, which match well according to the distance from Bolivia to United States and typical values obtained in South America.
We also observe that the IQR for IXP routes is smaller than for P2P and International routes, revealing a good performance of the IXP. It is worth noting that all links to the IXP are underused (the maximum percentage of occupation is $10\%$).

\begin{figure}[t]
\includegraphics[width=\columnwidth]{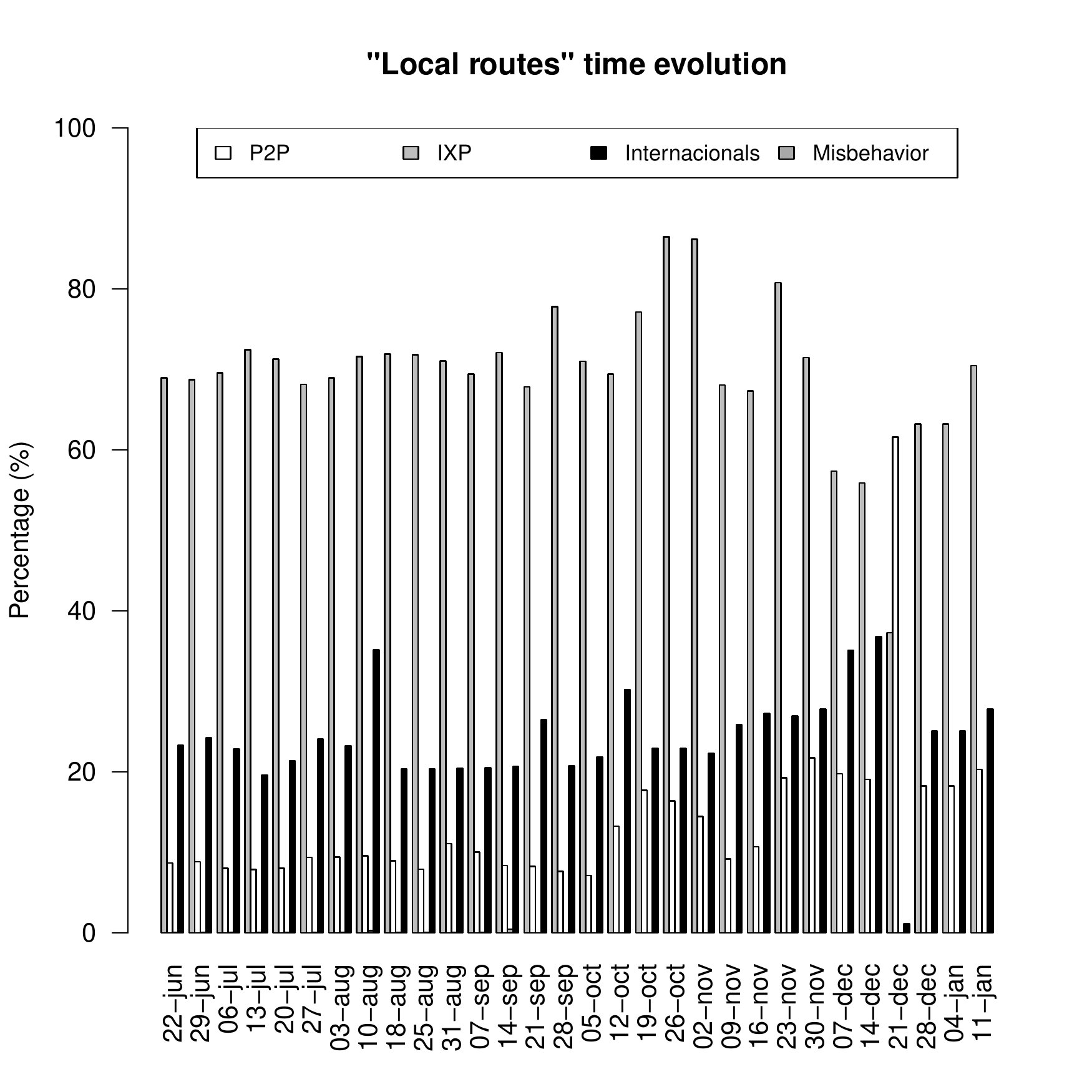}
\caption{Temporal evolution of local routes: percentage of {\tt traceroutes}  passing by each category, cumulated in a week.
}\label{local_routes}
\end{figure}

\begin{figure}[t]
\includegraphics[width=\columnwidth]{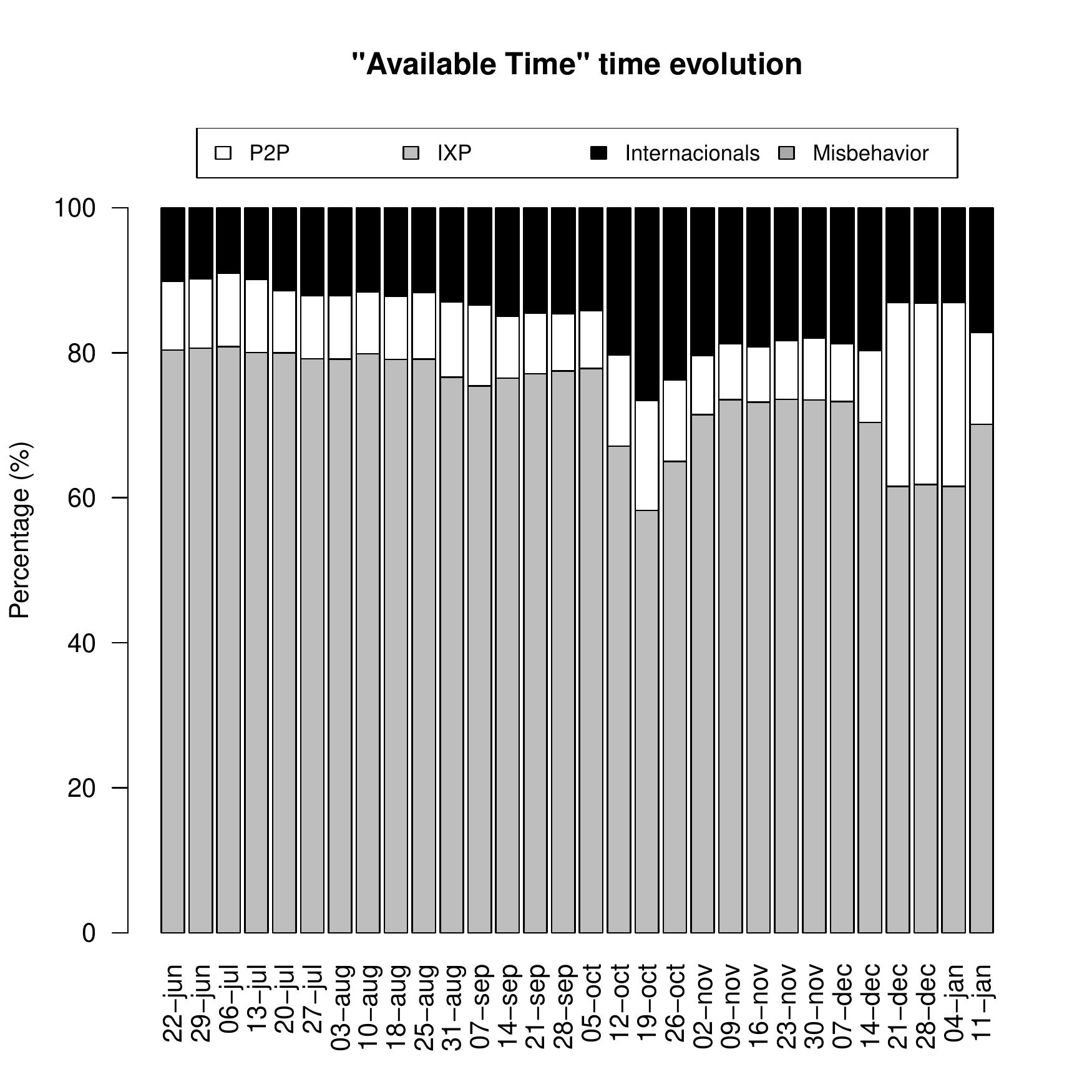}
\caption{Temporal evolution of available time: percentage of {\em routes} passing by each category, cumulated in a week.
}\label{used_time}
\end{figure}

The boxplots of temporal evolution of RTT are displayed in Figure~\ref{resu_rtt_t}. 
We continue using  95 and 5 percentiles for maximum and minimum.
Measurements in La Paz are stable, with a very low IQR for IXP routes. 
The period corresponding to both probes start with some important variations (probably due to changes in operators' networks), tending to historical values later.
The last period, which corresponds to the Santa Cruz probe, presents greater median and IQR values than those from La Paz in general. 
The evolution shows stable values for La Paz, increasing values for Santa Cruz in IXP and International only, but the tendency is to stabilize closer to previous values, mainly in the IXP category.
The measurements in La Paz exhibit a very low IQR for IXP routes, while in Santa Cruz are larger probably due to access through a 4G-LTE network. The same is observed for P2P routes. Notice that International routes seem to have similar IQRs than others, but due to the logarithmic scale they were larger as they was expected. 
This figures shows the bias introduced by the mean: for instance, the graph corresponding to International routes has a peak in 14-Sep and 21-Sep, where the maximum is more than 45 seconds (but 95 percentile is drawn); probably some problem arouse with the international connection of the Internet provider of the probe in La Paz. 
However, in 18-Aug (the last day of the analyzed week) there is a peak of the mean in the three categories that surprisingly match the 512k-day\footnote{The 12th August 2014 was the day when the BGP routing tables exceeded the 512k entries, causing connectivity problems in some widely used routers in Internet.}; also enforced by the 95-percentile increment.
This fact shows that even in the Bolivian network this effect was also present.

The next parameter is {\em local routes},  which seeks to measure the percentage of routes passing through the IXP, P2P or International links. 
Results are shown over one week, and they are computed taking the number of routes  (identified by a source and a destination addresses) crossing  IXP, P2P or International links, over the total number of routes. 
Notice that, as a certain route can pass by different type of links in different times throughout the week, the sum of the categories can be more than $100\%$. A good indication that the IXP is healthy is that the percentage of IXP routes is the highest one and International is the lowest one. 
Figure~\ref{local_routes} illustrates this parameter as a function of time. 
Notice that the IXP average is $70\%$, while International is $24\%$ and $14\%$ for P2P. 
The average value of routes measured per week is $1,753$, and in the week of 21-dec it is only $177$ which presents some non-consistencies with other weeks.

We also defined the {\em available time} parameter, which stands for the percentage of time that routes pass through either IXP, P2P or International links. 
It is computed by counting the number of {\tt traceroutes} passing through IXP, P2P or International links, over the total number of {\tt traceroutes}. 
Notice that a {\tt traceroute} is just an instance of a route: source and destination addresses. In this case the sum is always $100\%$ and Figure~\ref{used_time} exhibits this parameter as a function of time, measured in weeks. 
The IXP oscillates between $80\%$ and $60\%$, whereas some {\tt traceroutes} used International routes between $10\%$ and $15\%$.    

\begin{figure}
\includegraphics[width=\columnwidth]{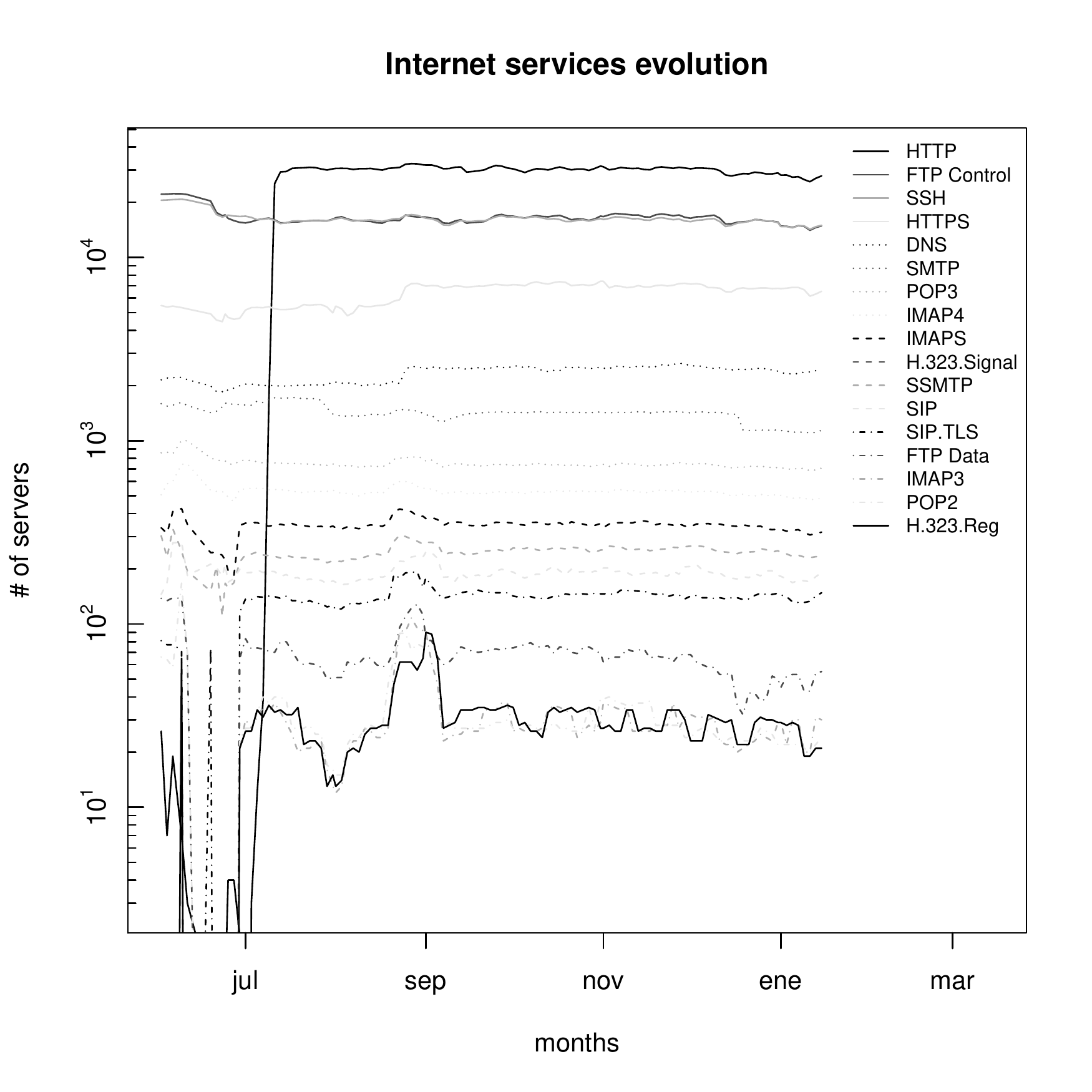}
\caption{Evolution of the amount of servers having IP addresses assigned to Bolivia, classified by TCP/UDP port.
}\label{ports}
\end{figure}

Figure~\ref{ports} shows the number of host offering different services with Bolivia-assigned IP addresses. 
The most important service is {\tt http}, followed by {\tt ftp} and {\tt ssh}, then the e-mail protocols and at bottom VoIP, streaming and secured e-mail protocols. 
There are a lot of variations in the beginning because there are some instabilities in our measurements, but from  late July they are stable. 
The general tendency is stable, and we believe it is related to the start of the IXP (operative since December 2013); it is needed certain time in the beginning to start developing  local services and to see increasing those in this graph. 
Furthermore, any CDN or a local DNS k-root mirror have not arrived to Bolivia yet.
This low increment, even HTTP ports, during this period motivated us to analyze which are the most visited websites in 
We got from Alexa database which are the 200 most accessed sites from Bolivia.
Once we collected this data, we analyzed which are using Bolivian IP addresses.
This processing is not accurate with popular sites, such as Google or Facebook, which can be placed locally though using foreign address.
However, no one of these sites have a server in Bolivia.
We found only 10 of these websites are using local addresses, nonetheless we expect that local hosting will be developed soon many of them will be hosted locally.

\begin{figure}
\includegraphics[width=\columnwidth]{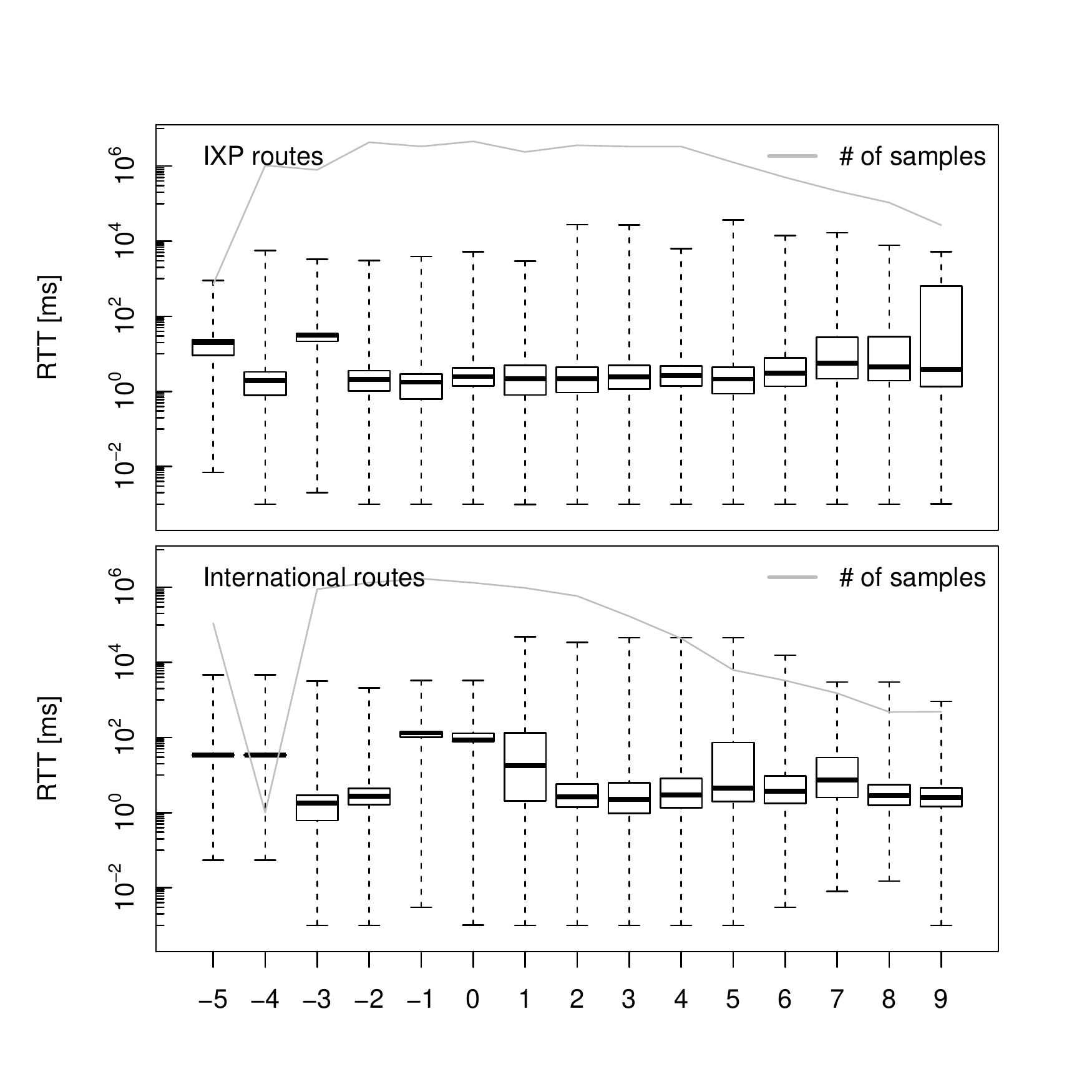}
\caption{Box-plots on the RTT difference between hops, centered at the IXP. 
The gray line shows the number of samples for each box-plot. 
Only routes passing through IXP links and positive results are considered.
}\label{diff_times}
\end{figure}

We next present RTT difference between adjacent hops in Figure~\ref{diff_times}, as a function of the distance to the IXP (the actual Bolivian IXP is a switch, therefore there are no internal hops).  
This RTT versus hop distance graph is similar to the one used by Hyun~\cite{hyun}.
Due to each hop, in a certain {\em traceroute}, it is done in a different time (usually one per second), so the RTT difference can give negative values based on several random phenomenons, none of these negatives values are considered.
Some of them are congestion queues, CPU processing time on routers and NAT devices, and asymmetrical paths between original packet and ICMP response.
When the difference of RTT of two consecutive hops is positive, it means these RTT are affected in a fairly similar way (e.g., congestion and processing). Therefore, we performed an statistical analysis for each hop and all traceroutes, in order to unveil the behavior of each hop at different distance to the IXP.


Previous works, such as  \cite{downey1999using} and \cite{jacobson1997pathchar}, have studied inter-hop relationship, mainly focused on bandwidth calculation. These works have used a delay time model, in which they could get deterministic parameters, in a congestion-free scenario, through minimum values. 
However, we found that minimum values for route is largely variable up on time-window sampling.
Instead of minimum values technique, we preferred to agreggate different routes at same hop-distance from the IXP.
Moreover, randomness given by reverse path variation may be mitigated using Reverse traceroute~\cite{katz2010reverse}. 
Nevertheless, this tool is based on IP header information, such as timestamp and route record, options which are disable in most of the Bolivian routers.


Figure~\ref{diff_times} presents boxplots, in logarithmic scale, displaying just the positive values of the differences. 
We inspected the shapes of the empiric density distributions and found that they look like lognormal ones; therefore IQRs are representative of the deviations. 
Notice that various minimum values are $1\mu s$, which is the time resolution given by operating systems. 
We also display the number of cases in each box-plot as a gray line. 
Hop zero corresponds to the IXP, and negatives are the ones previous to IXP while positives are the predecessors hops.  
The most representative samples are from $-5$ to $9$ (according to the number of cases),  which presents a comparative values of hops' medians, but in hop $-3$. 
We noticed that hop 0 is related to the IXP (remember that it consists in a switch) presenting a low IQR (around of $0.5ms$). 
Considering that RTT is related to the time expended in queues and physical transit, and assuming the same degree of symmetry for routes (i.e., the ICMP packet of TTL  exceeded follows the same route that the traceroute packet), the median can be related with distance of hops and IQR to the performance at that hop. For instance, from hops -4 to -3 there is big difference between medians, probably due to a long physical link (Bolivia is crossed by the Andes Mountains). From hop $-2$ to $5$ IQR remains fairly constant, which means that random phenomenons are not significant on the core of the Bolivian network. From $6$ to $9$ we can see how the IQR enlarges hop by hop. This may be related with the route aggregation that we have done. It is probably that on the last hops of the traceroutes, they reached a large number of different routers, which is shown on these increments.


In the Figure~\ref{diff_times} we also show, the RTT-differences for International routes. 
In this case, hop $0$ summarizes all the foreign links.

International routes hop $0$ has a median value of $100 ms$ approximately, and matches well with the journey done by the packets to reach the US four times.
Hop $-1$ represents the ASes border gateway, and it shows a very large inter-hop RTT, even greater than hop $0$.
This router might answer ICMP packets with large delay because is involved in BGP tasks, however, it does not explain this high value.
Another cause could be this border router is overloaded or it is an antique device.
Hop $1$ represents all first Bolivian routers after the internatinal journey of the packets, showing a large IQR.
The reasons of this large IQR could be at least two. 
First, the agreggation of multiple routes coming back to Bolivia; and second, a congestion queues on routers.
The other hops are placed in Bolivia with a similar behavior to IXP routes, though, with less amount of samples.

\begin{figure}
\includegraphics[width=\columnwidth]{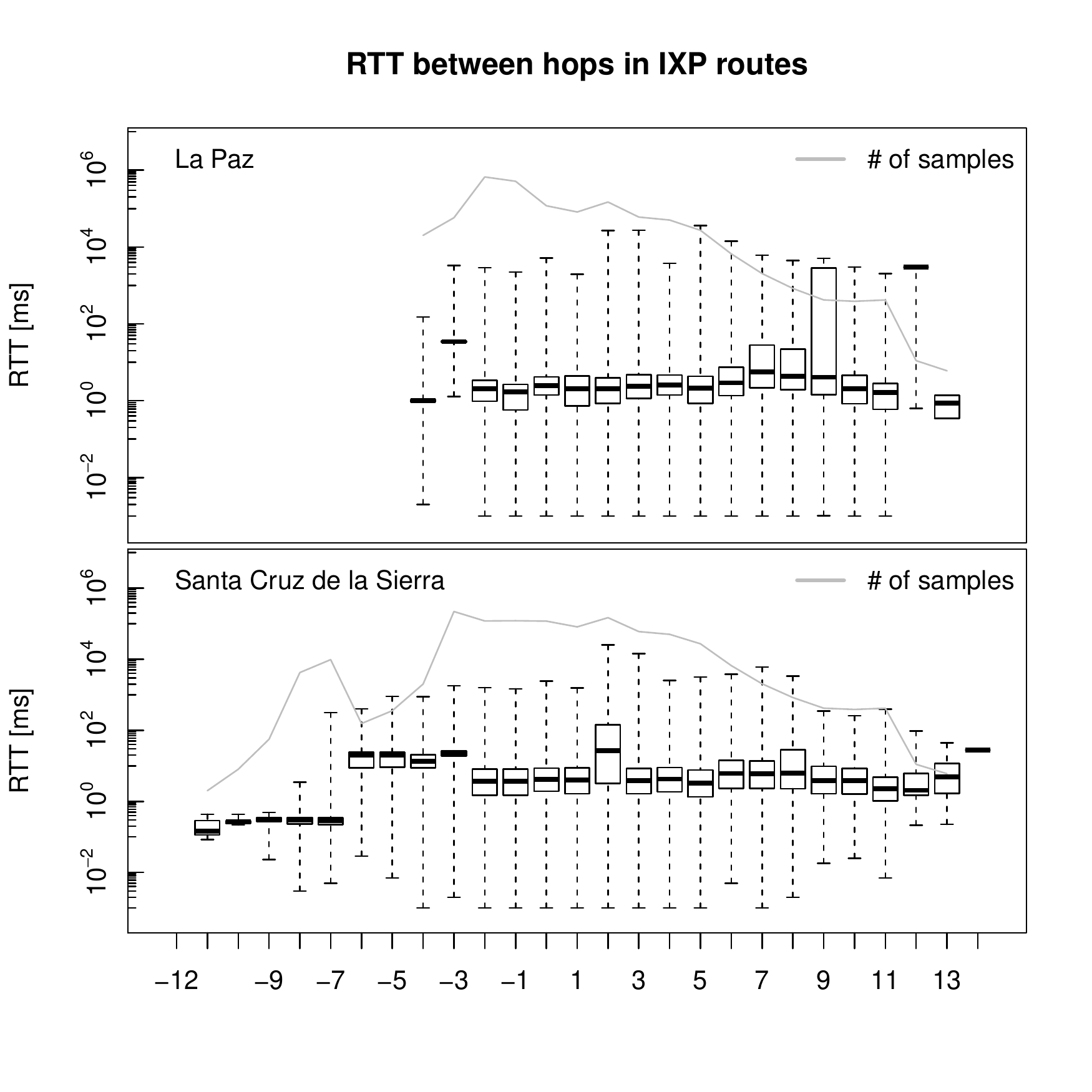}
\caption{RTT difference between hops for each probe, centered at the IXP. 
The gray line shows the number of samples for each box-plot. 
Only routes passing through IXP links and positive results are considered, and classified by probe.
}\label{diff_times_city}
\end{figure}

We also analyze the dependence of  RTT difference between adjacent hops as a function of the site probes in Figure~\ref{diff_times_city}. 
The first observation is that it is possible to distinguish that the low performance in hop 9 is due to routes from La Paz. 
Then, previous hops farther than hop -4 have no positive measurements or answers in {\tt traceroutes} from La Paz; we verified that most routes form this probe are at least at 6 hops from the IXP. 
Next, the  significantly difference between median of -4 and -3 hops is also from La Paz measurements.  
Following observation is related to hop 2 in Santa Cruz, where the performance is poor: the IQR is large.
Finally, Santa Cruz presents lots of hops before the IXP, which is concordant with the access technology that is 4G-LTE and the IXP is placed in La Paz.

\begin{figure}
\includegraphics[width=\columnwidth]{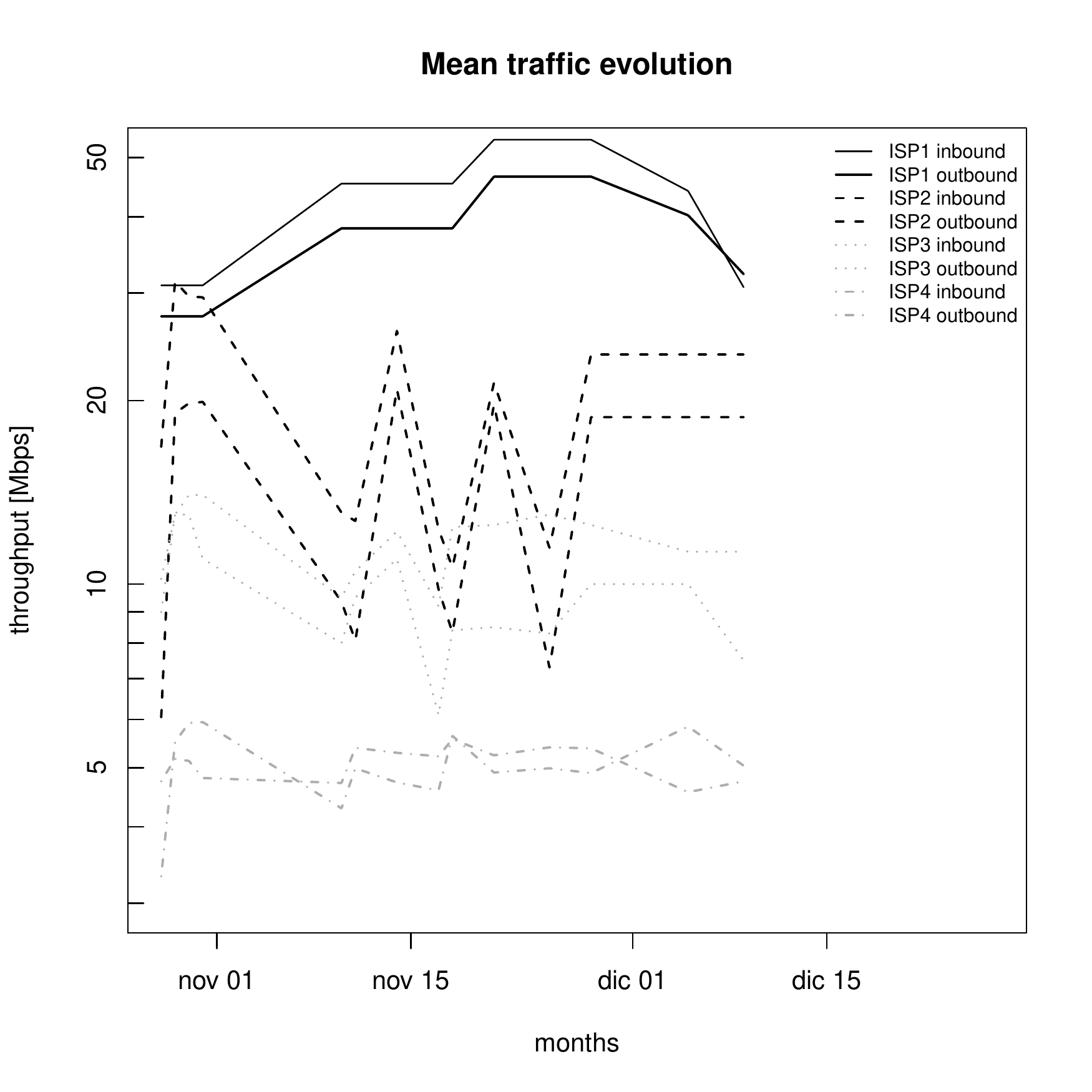}
\caption{Traffic evolution of some of the IXP members.
}\label{traffic}
\end{figure}

 We illustrate the evolution of traffic through the IXP of some of the ISPs connected to, in Figure~\ref{traffic}. 
 PladMeD is capable to obtain port utilization of the IXP switch using SNMP. However, IXP members are not allowed to share PlaMeD's SNMP access, yet. 
 Currently, some ISP are sharing their own IXP traffic statistics.   
 This shows  that we are on an early stage where there is not a progress, which is also confirmed by the stability of services analysis in Figure~\ref{ports}. 
 It is also remarkable the symmetry between in-bound and out-bound traffic.
 
\section{Conclusions}~\label{conclusion}

In this paper we present an analysis of the performance of the new Bolivian IXP.
Based on the taxonomy of IXP analysis papers (see Table~\ref{taxo_ixps}) we developed our tool called PladMeD, which provides flexibility to implement different measurement techniques.
We provide the image to install directly to the Rasberry PI, and a toy database which is already initialized to start measurements~\cite{PladMeD_RbPI}.        

We measure the evolution of several parameters of interest, in order to evaluate how the IXP performs and its impact on Bolivian network.
Key parameters are local routes, available time, operative servers and inter-hop RTT-difference. 
We analyze data for a 7-month period (June 2014 to January 2015) and examine how measurements from different probes can be merged to obtain more robust results.


This work has made a comparison of Internationals routes and IXP, showing improvements that IXP in Bolivia provides.
Due the geographical placement of Bolivia, the distance to Internet's backbone is large and these improvements are expected.
However, we can also show the percentage of route transiting through international links, allowing to understand the evolution and status of the ASes national interconnection.
As consequence of this research, we have seen that IXP improvements have not produced significant changes on local traffic o services availability yet. 
We expected that IXP's emergence on local networks would increase the number of subscribers, services and traffic as soon as they could.
However, the absence of CDNs, DNS k-root mirror and local hosting has delayed this increment, though, IXP has set a background for them.
Therefore, we saw that if the whole local Internet environment does not follow the changes proposed, setting local interconnection of ASes by  law only introduce little improvements on end-users experience. As these advantages are perceived by network operators as well as service providers in Bolivia, we expect traffic at the IXP to grow steadily.

Our future work is oriented towards adding more features to PladMeD, in order to better understand how traffic performs across hops and over different types of routes.

We also provide the inter-hop statistical analysis, which provides a first approach to understand the performance of physical network. Link distance affects the median value while congestion diversity of route affects the IQR. This new perspective, can provide baseline to compare with instantaneous measurements, i.e. an average of decene of traceroute can display if certain hop has congestion when is over a threshold (e.g., third quarter).

\section{Acknowledgments}
We acknowledge the support of Internet Society and ATT-Bolivia to this project. The authors also acknowledge the support of their institutions. 

%
 \bibliographystyle{unsrt}

 \bibliography{biblio}
%
%

\end{document}